\documentclass[aip,pop,reprint,amsmath,amssymb]{revtex4-1}

\usepackage{graphicx}
\usepackage{dcolumn}
\usepackage{bm}
\usepackage{float}
\usepackage{multirow}
\usepackage{color}

\begin{document}

\title{Molecular Dynamics Simulation of Strong Shock Waves Propagating in Dense Deuterium With the Effect of Excited Electrons}

\author{Hao Liu}
\affiliation{HEDPS, Center for Applied Physics and Technology, College of Engineering, Peking University, Beijing 100871, China}

\author{Yin Zhang}
\affiliation{Department of Mechanics and Engineering Science, College of Engineering,  Peking University, Beijing 100871, China}

\author{Wei Kang}
\email{weikang@pku.edu.cn}
\affiliation{HEDPS, Center for Applied Physics and Technology, College of Engineering, Peking University, Beijing 100871, China}

\author{Ping Zhang}
\affiliation{HEDPS, Center for Applied Physics and Technology, College of Engineering, Peking University, Beijing 100871, China}
\affiliation{ Institute of Applied Physics and Computational Mathematics, Beijing 100088, China}

\author{Huiling Duan}
\email{hlduan@pku.edu.cn}
\affiliation{HEDPS, Center for Applied Physics and Technology,  College of Engineering, Peking University, Beijing 100871, China}
\affiliation{Department of Mechanics and Engineering Science, College of Engineering,  Peking University, Beijing 100871, China}

\author{X. T. He}
\email{xthe@iapcm.ac.cn}
\affiliation{HEDPS, Center for Applied Physics and Technology,  College of Engineering, Peking University, Beijing 100871, China}
\affiliation{ Institute of Applied Physics and Computational Mathematics, Beijing 100088, China}
\affiliation{ IFSA Collaborative Innovation Center of MoE, Peking University, Beijing 100871, China }

\date{\today}

\begin{abstract}
We present a molecular dynamics simulation of shock waves propagating in dense deuterium with the electron force field method [J. T. Su and W. A. Goddard, Phys. Rev. Lett. 99, 185003 (2007)], which explicitly takes the excitation of electrons into consideration. 
Non-equilibrium features associated with the excitation of electrons are systematically investigated.
We show that chemical bonds in D$_2$ molecules lead to a more complicated shock wave structure near the shock front, compared with the results of classical molecular dynamics simulation. 
Charge separation can bring about accumulation of net charges on the large scale, instead of the formation of a localized dipole layer, which might cause extra energy for the shock wave to propagate.
In addition, the simulations also display that molecular dissociation at the shock front is the major factor corresponding to the ``bump'' structure in the principal Hugoniot. 
These results could help to build a more realistic picture of shock wave propagation in fuel materials commonly used in the inertial confinement fusion.
\end{abstract}

\maketitle


\section{INTRODUCTION}

Inertial confinement fusion (ICF) \cite{Nakai1996, Lindl1995PP1} is an effective way to generate energy.
The method requires to compress the fusion fuel, which is composed of hydrogen isotopes deuterium (D) and tritium (T), into an extreme state over 1,000 times of the solid density. 
This high-density condition is difficult to attain through static compressions with current techniques. 
Dynamical approaches are commonly used instead, in which shock waves driven by high-power lasers or explosives are employed to compress the fuel. 
Therefore, understanding how strong shock waves compress the fuel  is a necessity in the study of ICF.

Much effort has been devoted to understanding the structure of strong shock waves in various materials \cite{Zeldovich1965,Muckenfuss1962PF1, Klimenko1978, Holian1979, Holian1995,Hoover1979, Hoover2009, Zhakhovskii1997, Zhakhovskii1999,Lin2014}. 
Early theoretical works based on the Navier-Stokes equations\cite{Zeldovich1965,Hoover1979} and the Boltzmann equation \cite{Mott-Smith1951,Muckenfuss1962PF1} provide a basic physical picture of shock waves in fluids.
With the development of computational techniques, in particular with the advent of efficient numerical hydrodynamic methods and codes \cite{Ramis1988CPC,Harte1996FWTCHIPDHF}, it is possible to show the structure of shock waves with increasing details and under complicate conditions close to those experienced by the fuel material.
For strong shock waves, the underlying local thermal equilibrium (LTE) assumption in theoretical investigations can be partially removed by the classical molecular dynamics (MD) method \cite{Klimenko1978,Holian1995,Liu2016fop,Zhakhovskii1997}. 
It takes kinetic effects of atoms into account, and thus gives a better description of the shock wave structure.  
Classical MD simulations performed at various levels of technical sophistication \cite{Holian2010jcp, Hoover2009, Zhakhovskii1997, Belonoshko2005} have shown that there are highly non-equilibrium behaviors, including shock induced phase transitions \cite{Fortov2007PRL,Collins1998S}, and molecular dissociation \cite{Belonoshko2005, Chen2010PL}, in the region near the shock front. These findings have stimulated further development of shock wave theories \cite{Garcia-Colin2008,Lin2014, Holian1993, Holian2010jcp, Holian2010pre}.

Unlike in weak shock compressions, where the material properties are mainly determined by degenerate electrons, it has been well recognized \cite{Zeldovich1965,Su2007PRL, Liberman1986SVSSE} that the excitation of electrons is an essential factor that has to be taken into account in the compression of fuel materials. 
The excitation appears as, for example, strong ionization and charge separation near the shock front at high impact velocity.
It becomes significant when the downstream temperature of the shock wave rises to above 5 eV, which is typical in the implosion process of ICF. 

Although there are a few of methods \cite{Marques2004ARPC, Militzer2000PRL, Mermin1965PR} that can take the excitation of electrons into consideration, the actual choice of methodology is quite limited as long as the non-equilibrium feature of both electrons and ions is concerned. 
Methods that explicitly depend on the electronic temperature, e.g.,  the finite-temperature density functional theory (time-independent version) \cite{Mermin1965PR}, have to be excluded from the list, because local temperature has been demonstrated in previous works \cite{Liu2016fop,Holian1993,Zhakhovskii1997} not well defined in the highly non-equilibrium region near the shock front.

The non-equilibrium feature of electrons can be captured when the time-dependent dynamics of electrons is included faithfully.
However, preceding attempts to do this on the level of the time-dependent density functional theory (TD-DFT) \cite{Kang2016SR,  Zhao2015PL,Baczewski2016PRL}  have shown that this approach is extremely computationally costly.
Practically, it is only capable to include tens of atoms in the calculation, which is far less than the required number of atoms to describe the propagation of shock waves.
So, it is more realistic to use some simplified version of time-dependent electronic dynamics, e.g., the so called wave-packet molecular dynamics (WPMD) method \cite{Knaup2001CPP, Knaup2003JPMG, Su2007PRL, Graziani2014}, which approximates electronic wave functions as Gaussian wave packets and describe the dynamics of electrons through the average position and smearing (size) of the wave packets. 

In this work, the excitation of electrons is described by the electron force field (eFF) method \cite{Su2007,Su2007PRL,Su2009jcp,Jaramillo-Botero2011}, which is a further development of the WPMD method. 
In addition to the Gaussian wave packet approximation to electronic wave functions, the eFF method provides a simplified parameterization with improved accuracy to the Pauli's exclusion force between electrons of the same spin, which is the necessary part in the description of electronic structures, e.g., the shell structure and chemical bonds. Further studies of fuel materials in the equilibrium states \cite{Su2007,Su2007PRL} have shown that this method can also be applied to materials under high energy density conditions \cite{Drake2006} typical in the ICF experiments. This encourages us to employ it in the investigation of dynamical structures of shock front.

We present a molecular dynamics simulation of shock waves propagating in deuterium with the eFF method, where non-equilibrium features associated with the excitation of electrons are addressed.
We show with the simulation that chemical bonds in D$_2$ molecules lead to a more complicated shock wave structure near the shock front, and charge separation at the shock front brings about accumulation of net charges on the large scale, instead of the formation of a localized dipole layer, which may cause extra energy for the shock wave to propagate.
In addition, the simulation also displays that molecular dissociation at the shock front is the major factor corresponding to the ``bump'' structure in the principal Hugoniot.

The rest of the article is organized as follows. Theoretical description of the eFF method and computational details are presented in Sec.~\ref{sec_2}.  Sec.~\ref{sec_3} is the main results and discussions. Then a short summary in Sec.~\ref{sec_4} concludes the entire work.

\section{Methodology and Computational Details}\label{sec_2}

The propagation of shock waves in deuterium is simulated with a combination of classical molecular dynamics method for ions and the eFF method for electrons. The interaction  between ions and electrons are assumed to be adiabatic and forces between them are calculated through the Ehrenfest's theorem \cite{Liboff1980A}.
Electron-electron interaction is described by the eFF method proposed by Su and Goddard \cite{Su2007PRL}. 
In the eFF method, the electronic wave function is approximately described by a Gaussian wave packet 
\begin{equation}
\nonumber
\Psi({\bf r}) \propto \prod_{j} \exp\left\{-\left[\frac{1}{s^2}-\frac{2p_{s}}{s}\frac{i}{\hbar}\right]({\bf r}-{\bf x})^2-\frac{i}{\hbar} {\bf p}_{\bf x}\cdot {\bf r}\right\},
\end{equation}
where ${s}$ represents the smearing (the size) of the wave packet, {\bf x} is the average position of the wave packet,  and $p_s$ is the conjugate momentum of $s$.
The semi-classical equations of motion for $\bf x$ and $s$ are derived \cite{Heller1975} by inserting the wave packet approximation into the time-dependent Schr\"{o}dinger equation, which leads to
\begin{equation}
\nonumber
\begin{split}
\dot{{\bf p}_{\bf x}}&=-\frac{\partial V}{\partial {\bf x}},  \qquad  {\bf p}_{\bf x}=m_e \dot{\bf x}, \\
\dot{p_{s}}&=-\frac{\partial{V}}{\partial{s}},  \qquad p_{s}=\frac{3}{4} m_{e} \dot{s}, \\
\end{split}
\end{equation}
with $V=V_{ii}+V_{ie}+V_{ee}+E_{_{KE}}+E_{_{PR}}$. Here, $V_{ii}$, $V_{ee}$ and $V_{ie}$ represent the ion-ion, electron-electron and ion-electron interactions respectively. $E_{_{KE}}$ and $E_{_{PR}}$ are the kinetic energy of the Gaussian wave packet and Pauli repulsion energy, which account for quantum mechanical effects of electrons. 
There are several sophisticated constructions for the expression of $E_{_{PR}}$. In our calculation, we use the simplest one following Ref.~\onlinecite{Su2007}.

All simulations are performed using the eFF implementation included in the molecular dynamics code LAMMPS\cite{Plimpton2007}.
The electron mass is set as $m_{e}=0.01$ amu to perform the simulation with a relatively large time step $\Delta t=0.01$ fs. 
Simulations with $m_{e}=0.1$ are also carried out to illustrate the mass effect of electrons.
Note that the $m_e$ here is different from that used in Su and Goddard's original work, in which $m_e$ is set as 1 amu \cite{Su2007}. 
It is reasonable to set $m_e$ = 1 amu for systems at equilibrium, but in a dynamical simulation, it would be better to give electrons a smaller mass to capture the charge separation effect.
 
The simulation box has a size of 102.271 Bohr $\times$ 102.271 Bohr $\times$ 33 749.5 Bohr, corresponding to the length along the x, y and z axes. 
Initially, the simulation box is filled with 2 640 000 deuterium atoms and 2 640 000 electrons. 
The initial Wigner-Seitz radius of deuterium atoms is $r_s = 3.1722$ Bohr, corresponding to $\rho_0$ = 0.169 g/cc. 
Before shock waves propagate, the entire system is thermalized to a liquid state of deuterium at $T_0=$ 20 K and $P_0=$ 27 MPa.

\begin{table}
\caption{\label{table1} Shock wave parameters extracted from the MD simulations, where $v_p$ is the piston velocity, $v_s$ is the shock wave speed, and $\eta$ is the compression ratio. $T$ and $P$ are temperature and total pressure in the downstream region of the shock wave.}
\begin{ruledtabular}
\begin{tabular}{ccccc}
$v_p$&
$v_s$&
$\eta$&
$T$&
$P$\\
\textrm{(km/s)}&
\textrm{(km/s)}&
\textrm{}&
\textrm{(K)}&
\textrm{(GPa)}\\
\colrule
\\
20&	25.2&	4.8&	6 600&	89\\
30&	37.2&	5.2&	13 000&	190\\
40&	49.8&	5.1&	25 000&	330\\
50&	62.7&	4.9&	68 000& 	510\\
70&	88.5&	4.8&	140 000& 1000\\
\\
\end{tabular}
\end{ruledtabular}
\end{table}

Periodic boundary conditions along the x and y axis are assumed throughout the simulation.
A reflective wall moving at a constant speed $v_p$ is used as the piston to drive the shock wave.
The piston is placed at one end of the z axis so that the shock wave travels along the positive $z$ direction.
At the other end of the z axis, a fixed reflective wall is used to keep the deuterium atoms in the simulation box. 
Simulations will be terminated before fast electrons hit the reflective wall in order to remove its influence on the shock wave structure.
The piston speed $v_p$ varies from 20 km/s to 70 km/s. The corresponding  shock velocity $v_s$ is ranged from 25.2 km/s to 88.5 km/s,  as summarized in Table~\ref{table1}.

The cutoff for pair interactions is 10 Bohr, which is more than 3 times of the Wigner-Seitz radius of deuterium atoms.
It takes all the interactions of the nearest and the next-nearest neighbors into consideration. 
The value of the cutoff is a trade off between computational efficiency and the size of simulation. 
Neglecting the long range part of the Coulomb interaction will lead to an overestimation to charge separation, but will not change the qualitative physical picture.

Profiles of macroscopic flow variables, such as temperature, density, and electrical field,  are calculated in the coordinate systems moving with the shock front. 
Their values presented in the work are the spatial average in small slices of 4 Bohr thickness along the z axis \cite{Liu2016fop}.

A quantity that one needs to pay special attention to is the electronic temperature $T_e$, which is derived from the wave packet form of the wave function. 
In principle, it is defined as \cite{Su2007}
\begin{equation}\label{eq 1.3}
T_e=\frac{1}{4Nk_B} \sum_{\alpha}^N  m_{e} (\mathbf{v}_{x,\alpha}^{2}+ \frac{3}{4}v_{s,\alpha}^2), 
\end{equation}
where $N$ is the number of electrons in the calculation slice, and $k_B$ is the Boltzmann constant. $\mathbf{v}_x$ and $v_s$ represent $\dot{\mathbf{x}}$ and $\dot{s}$ respectively.
The subscript $\alpha$ denotes the $\alpha$-th electron in the calculation slice. 
The secend part in Eq.~(\ref{eq 1.3}) is the xxx xxxx.
Note that $T_e$ approaches the real electronic temperature only at high temperature.
At low temperature, e.g., in the initial state, when most electrons are in bonded states, $T_e$ calculated from Eq.~(\ref{eq 1.3}) will essentially deviate from the real value, and thus can not be interpreted quantitatively.  

\section{Results and Discussions}\label{sec_3}

\subsection{Structure of shock fronts}\label{subsec_hydro}

\begin{figure}[]
\includegraphics[width=0.45\textwidth]{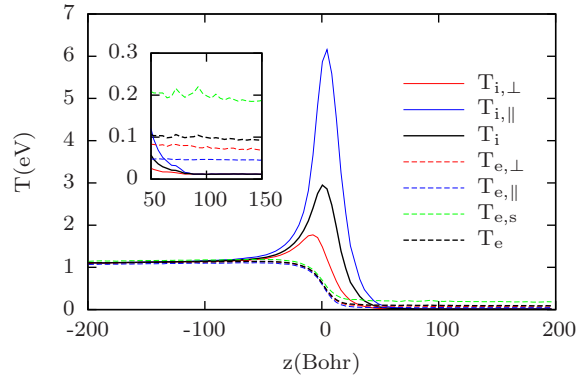}\\
\caption{Profiles of temperature components along the z axis for $v_p$ = 30 km/s. They represent typical temperature distributions of strong shock waves. The inset is the zoom-in of temperature components in the upstream region.}
\label{pic_temperature}
\end{figure}

With the eFF method, one can have an atomic resolution for the fine structure of shock fronts, which, by taking electrons into consideration, displays different features compared with those revealed by classical MD method \cite{Liu2016fop,Holian1993,Zhakhovskii1997} or by other non-equilibrium methods \cite{Garcia-Colin2008,Yen1966PF,Lin2014} that do not take electronic excitation into consideration.

An important feature of the shock front structure is the strong ``overshoot''\cite{Yen1966PF}, i.e., a high peak, of ion temperature and its components near the shock front, which is much weaker in the classical MD simulations \cite{Liu2016fop,Holian2010jcp,Zhakhovskii1997}.
Fig.~\ref{pic_temperature} shows the distributions of all temperature components near the shock front, including those of the ion temperature $T_i$, electron temperature $T_e$, as well as their components $T_{i,\parallel}$ , $T_{i, \perp}$, $T_{e,\parallel}$, $T_{e,\perp}$, and $T_s$.

\begin{figure}[]
\includegraphics[width=0.45\textwidth]{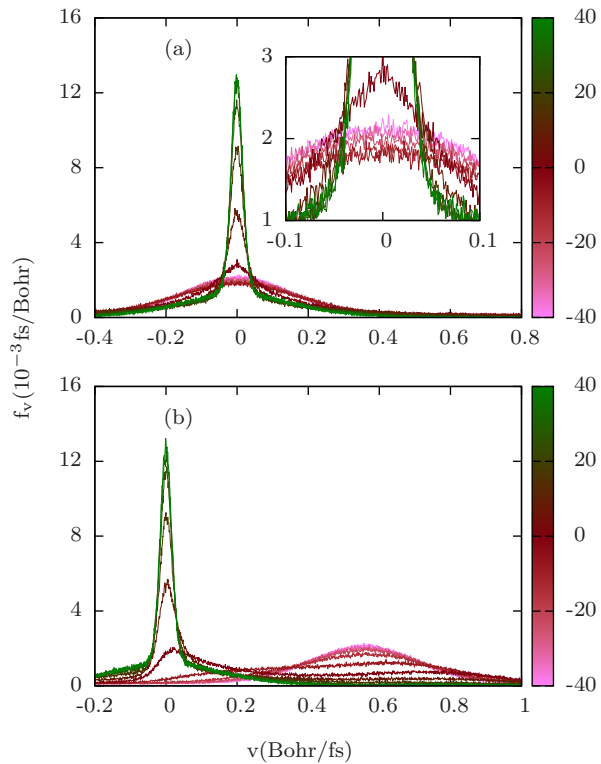}
\caption{Ion velocity distributions at selected positions near the shock front, in the simulation with a piston velocity $v_p$ = 30 km/s. The color of each curve represents the distance with respect to the center of the shock front, as denoted by the color bar on the right side. (a) Velocity distribution on the perpendicular direction, and (b) the same as (a) but on the parallel direction, with respect to the propagation direction of shock waves.}
\label{pic_vdis}
\end{figure}

The ion temperature and its components have a protruding high peak at the shock front. 
It is a typical non-equilibrium phenomenon associated with the relaxation of ions \cite{Zeldovich1965}.
Classical MD simulations have shown that \cite{Liu2016fop, Holian2010jcp, Zhakhovskii1997} only $T_{i,\parallel}$  in a one-component medium has a perceivable ``overshoot'' near the shock front.
However, Fig.~\ref{pic_temperature} shows that $T_{i,\perp}$  also displays a significant ``overshoot''  in addition to $T_{i,\parallel}$.
The peak of $T_{i,\parallel}$  is about 6 times of the $T_{i,\parallel}$'s value in the downstream region far from the shock front.
This ratio is much larger than that (1.5 times) observed in the classical MD simulations \cite{Liu2016fop}. 
It suggests that, in addition to the kinetic relaxation revealed by MD simulations \cite{Liu2016fop,Holian2010jcp,Zhakhovskii1997}, there is an extra relaxation process taking place on both the parallel and perpendicular directions (with respect to the traveling direction of shock wave.)
This extra relaxation process is attributed to the bond-breaking process of D-D bonds, as will be further discussed in Subsection \ref{subsec_effect}.

The transition of electron temperature and its various components at the shock front is much smoother than that of ion temperatures.
 No ``overshoot'' is observed in either of them. 
The difference between the distributions of $T_i$ and $T_e$ is originated from the much smaller mass of electrons (0.01 amu in the simulation) compared with that of a deuterium atom. 
Roughly speaking, the relaxation time of an ensemble of particles  is proportional to the square root of their mass, as estimated from the classical theory of plasmas \cite{Zeldovich1965}. 
This means that the relaxation process of electrons is about 10 times faster than that of ions, and thus difficult to observe in the transition region at a spatial resolution of 4 Bohr.

The upstream region of the shock front is enlarged in the inset of Fig.~\ref{pic_temperature}. It shows that the value of all components of $T_{e}$ in the upstream region is much higher than the components of $T_i$, which is 20 K in the simulation. 
This is not surprising since $T_e$ has a quantum-mechanical origin, and it should be aware when interpreting the data quantitatively. 

Velocity distribution of ions at various positions with respect to the shock front are displayed in Fig.~\ref{pic_vdis}.
The distribution of the $v_{\parallel}$ component is similar to that reveal by classical MD simulations \cite{Zhakhovskii1997, Liu2016fop}, whereas the distribution of $v_{\perp}$ shows a slightly different feature corresponding to the overshoot of $T_{i,\perp}$. 
As displayed in the inset of Fig. \ref{pic_vdis} (a), the height of peaks in the $v_{\perp}$ distribution  keeps increasing when the observing position in the downstream region leaves the shock front. In a Maxwellian velocity distribution, the increase of peak height corresponds to a decrease in temperature. This increasing trend displayed in $v_{\perp}$ thus corresponds to the drop of ion temperature at the rear of the shock front,
which is in line with the relaxation of the ``overshoot'' in $T_{i,\perp}$, as displayed in Fig.~\ref{pic_temperature}.

\subsection{Charge separation}\label{subsec_charge}

\begin{figure}[]
\includegraphics[width=0.45\textwidth]{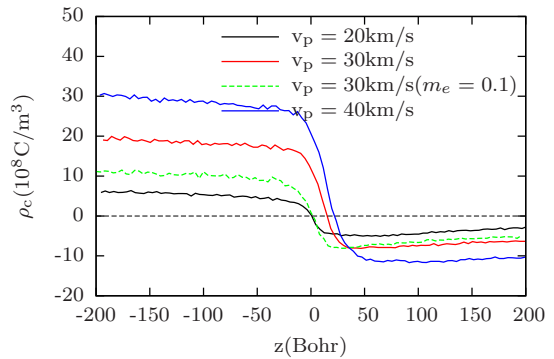}
\caption{Profiles of net charge density along the z axis for different piston velocities and electron masses. The black, red and blue solid lines correspond to $v_p =$  20, 30 and 40 km/s and $m_e$=0.01 amu, while the green dashed curve corresponds to  $v_p=$ 30 km/s and $m_e$=0.1 amu.}
\label{pic_charge}
\end{figure}

With the eFF method, charge separation at the shock front can be illustrated in the dynamical simulations.
When deuterium atoms are ionized, part of the bounded electrons become free electrons.
They have a larger translational thermal velocity than that of ions because $m_e \ll m_i$, and thus have a longer thermal diffusion length that can penetrate deeper into the upstream region (when observed in a reference framework moving with the shock front.)
When a considerable number of ionized electrons penetrate the shock front, which leave ions with positive charges behind in the downstream region, a region of non-vanishing net charge density emerges near the shock front.
As displayed in Fig.~\ref{pic_charge}, negative charges are concentrated in the upstream region and positive charges are in the downstream region.

The influence of $m_e$ can be further illustrated by setting $m_e$ to be 0.1 amu. As displayed by the green dashed curve in Fig.~\ref{pic_charge}, a significantly lower degree of charge separation can be observed, compared with the simulation with $m_e$ = 0.01 amu, while both have the same piston velocity of $v_p$ = 30 km/s.
The accumulation of net charges in the downstream region is in contrast to the traditional picture of charge separation near the shock front, in which a localized ion-electron dipole layer at the shock front is formed\cite{Zeldovich1965,Liberman1986SVSSE}, and the thickness of the dipole layer is on the same order of the shock front thickness.
The picture of localized charge separation is important to most of the radiative hydrodynamic programs \cite{Ramis1988CPC,Harte1996FWTCHIPDHF}, in which the charge separation is entirely neglected because its spatial extension is considered  much less than the resolution of the grids. 
Our results are quite unexpected at first glance.
It turns out resulted from the lacking of electron supplies at the downstream region of the shock front, where the reflective piston used in the simulation is impenetrable.

Whether the accumulation of net charges represents a real experimental situation depends on the setup of experiments. 
In gas gun experiments \cite{Holmes1995PR,Gu2009JCP}, the downstream flow can  get electron supplies from the environment, e.g., the wall of the container. 
There is no problem to maintain the charge neutrality on the large scale.
However, in a typical implosion experiment of ICF \cite{Lindl2004PP1}, where the fuel parcel is driven by X-ray radiations, the downstream flow of the shock wave does not get external electron supplies once fast ionized electrons move inward to the center of the fuel parcel, assuming no convective instability is intrigued by the strong electric field induced by the net charges. 

A direct consequence for the accumulation of net charges is that it costs extra energy, which decreases the energetic efficiency of the driver.
Since a relatively small cutoff of 10 Bohr is used to calculate the interaction between particles in our simulation, and the long-range part of the Coulomb interaction is neglected, Fig.~\ref{pic_charge} provides an overestimated accumulation, and  can thus only used as a qualitative demonstration.
A more accurate estimation of this effect in ICF is beyond the scope of current work. 
It might be done with the simulation techniques that include a faithful description to the excited electrons as well as to the coupling between radiative field and hot dense plasmas \cite{Drake2006}.

\subsection{Molecular dissociation and ionization at the shock front}\label{subsec_effect}
\begin{figure}[]
\includegraphics[width=0.45\textwidth]{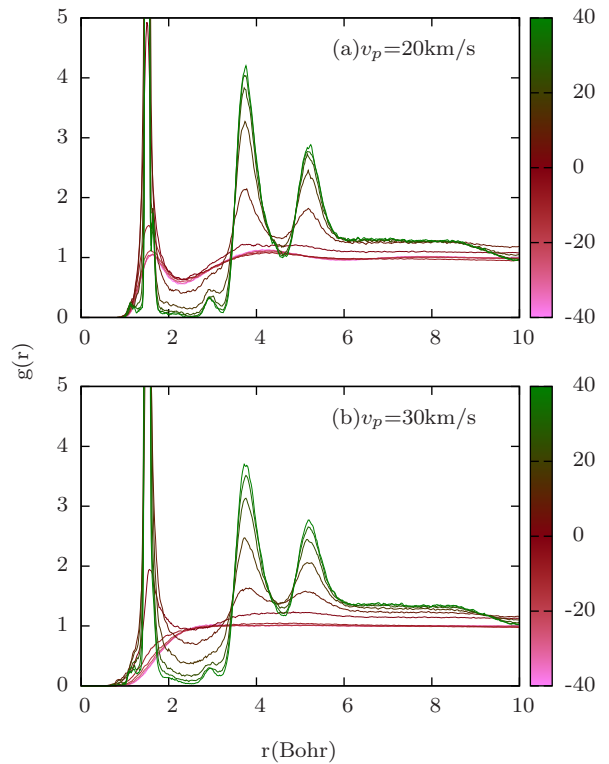}
\caption{Radial distribution functions at selected positions along the propagating direction of shock waves. The color of each curve corresponds to its distance from the center of the shock front, which is denoted in the color bar on the right side. Two cases are displayed corresponding to different piston velocities. In (a), $v_p$ = 20 km/s, and in (b) $v_p$ = 30 km/s. }
\label{pic_gr}
\end{figure}

With the eFF method, dissociation of chemical bonds is clearly displayed. Fig.~\ref{pic_gr} shows the radial distribution function (RDF) $g(r)$ of ions at various positions with respect to the shock front for $v_p$ = 20 km/s and $v_p$ = 30 km/s.
RDFs in front of the shock wave are presented as green curves, while those at the rear are displayed as pink lines. 
The first peak of the RDF in Fig.~\ref{pic_gr}, located at $r \sim$ 1.4 Bohr (0.74 \AA), corresponds to the D-D bond of D$_2$ molecules. 
The RDF also displays two additional  peaks at larger $r$ ($\sim$ 3.7 Bohr and $\sim$ 5.2 Bohr) in the upstream region of the shock front. They are attributed to the atoms of the nearest molecules. 
The peaks corresponding to the next nearest molecules disappear in the RDF, which indicates that the initial state has a liquid structure. The height of the peaks decreases along with the shock transition, showing that a phase transition takes place at the shock front.

The height of the first peak is also a qualitative measurement of molecular dissociation. 
For both cases displayed in Fig.~\ref{pic_gr}, the height of the first peak is significantly changed when the observing position crosses the shock front.
In addition to that, also observed is the broaden of the peak width resulted from the increase of temperature. 
At positions away from the shock front, the height of the first peak is nearly a constant, which suggests that the dissociation of D$_2$ molecules takes place in the transient region near the shock front, and is synchronized with the passage of the shock wave. 
Although fast ionized electrons arrive before the shock front, as illustrated by the charge density profile in Fig.~\ref{pic_charge}, they do not cause recognizable dissociation of D$_2$. 
This shows that the dissociation is essentially resulted from the kinetic effect of ion collisions. The impact of electrons has a small influence on the breaking of  D-D bonds. 

\begin{figure}[]
\includegraphics[width=0.45\textwidth]{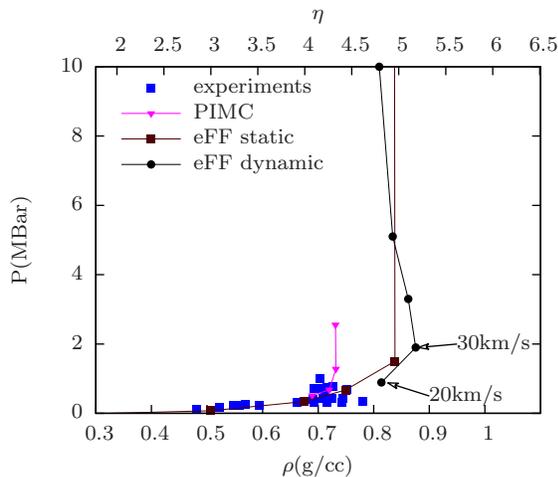}
\caption{Principal Hugoniot measured in the eFF ``dynamic" simulations of shock propagation, compared with those obtained from other approaches. The two arrows indicate the two states corresponding to $v_p$ = 20 km/s and $v_p$ = 30 km/s. Results from the eFF method through the Rankine-Hugoniot relation is taken from Ref.~\onlinecite{Su2007}, experimental results are taken from Refs.~\onlinecite{Holmes1995PR, Knudson2001PRL, Knudson2003PRL}, and the PIMC results are from Ref.~\onlinecite{ Militzer2000PRL}. }
\label{pic_hugoniot}
\end{figure}

The first peak of $g(r)$ in the downstream region disappears  between $v_p$ = 20 km/s and $v_p$ = 30 km/s, as can been seen by comparing Fig.~\ref{pic_gr} (a) and (b). 
These two states are also indicated with arrows in the principal Hugoniot in Fig.~\ref{pic_hugoniot}. 
It shows that these two states are located near the maximum compression ratio of the principal Hugoniot. 
This gives a strong support to the physical picture that the origin of the ``bump'' in the deuterium Hugoniot curve is the dissociation of D$_2$ molecules. 
Usually, the ``bump'' structure in the principal Hugoniot is the result of ionization of multi-shell electrons \cite{Zhang2016PP1, Rozsnyai2001PL}.
However,  it has a slightly different origin in the principal Hugoniot curve of D$_2$.

Ionization needs higher energy than molecular dissociation in most cases.
So, the ionization ratio is expected to be much smaller than that of the molecular dissociation. 
The number of ionized atoms in each calculation slice is estimated through 
$N_{\text{ionized}}=N_l+N_c,$
which has taken into consideration the fact that a portion of ionized electrons escape from the downstream to the upstream region of the shock front.
Here, $N_{\text{ionized}}$ is the number of ionized atoms, $N_l$ is the number of electrons of which the size parameter $s$ is larger than a threshold $r_c$.
$N_c = N_i-N_e$ is the number of net charges in the slice. 
The ionization ratio $\alpha$ is then calculated as $\alpha = N_{\text{ionized}}/N_i$.
In our calculations, $r_c$ is chosen as 50 Bohr , which is half of the length of the simulation box along the x and y axis, as suggested in Ref.~\onlinecite{Su2007}.
Note that the absolute value of $\alpha$ depends on the choice of $r_c$, and might not be the same as experimental measurements. 
It provides a reasonable physical picture for the ionization of deuterium under shock impact.

\begin{figure}[]
\includegraphics[width=0.48\textwidth]{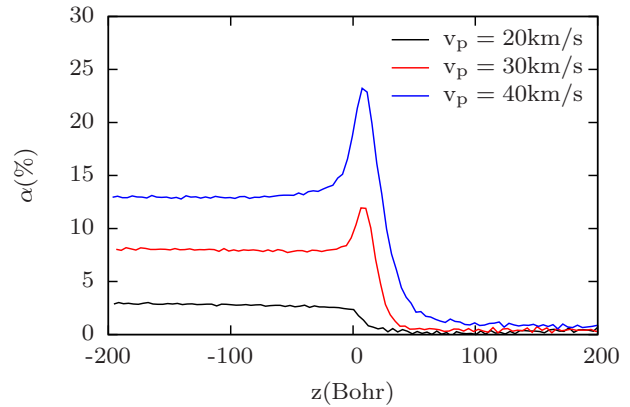}\\
\caption{Ionization ratio profiles along the propagating direction of shock waves. Profiles of different color correspond to different piston velocities, as indicated in the legend. }
\label{pic_ionization}
\end{figure}

The profiles of $\alpha$ for a variety of shock strengths are displayed in Fig.~\ref{pic_ionization}.
The average ionization ratios are 3\%, 8\% and 13\% corresponding to piston velocities of 20, 30 and 40 km/s. 
Note that at $v_p$ = 30 km/s, most of the D$_2$ molecules are dissociated, whereas only 8\% are ionized, which is a small fraction of the atoms. This shows that ionization is not a main resource for the ``bump'' structure in the principal Hugoniot curve of D$_2$.  
It is also noticed that there is a peak of $\alpha$ in the transition region of shock front, which corresponds to an ``overshoot'' of the ionization and its recovering process. 
The similarity of this peak structure with that of $T_i$  displayed in Fig.~\ref{pic_temperature} suggests that the ionization at the shock front is induced by the kinetics of ions.

\subsection{Principal Hugoniot from dynamical simulations}\label{subsec_hugoniot}

There have been extensive efforts \cite{Holmes1995PR,Knudson2001PRL, Knudson2003PRL,Da-Silva1997PRL,Collins1998S, Pierleoni1994prl, Militzer2000PRL,Lenosky2000PR} to measure and calculate the principal Hugoniot of D$_2$, which serves as a major benchmark for various equations of state (EOS) used in ICF.
Experimentally, the high energy density state is generated by driving a shock wave in the sample materials using gas guns\cite{Holmes1995PR,Chen2010PL}, exploding wires\cite{Knudson2001PRL, Knudson2003PRL} or lasers \cite{Da-Silva1997PRL,Collins1998S}, which is similar to the setup used in the dynamical simulations.
On the other hand, there are  a number of theoretical methods \cite{Pierleoni1994prl, Militzer2000PRL,Lenosky2000PR} which estimate the principal Hugoniot of D$_2$ via the Rankine-Hugoniot relation \cite{Zeldovich1965}.

In Fig.~\ref{pic_hugoniot}, the square dots represent the Hugoniot curve estimated from the dynamical simulations with the eFF method. 
It is compared with the Hugoniot calculated with the same method but through the Rankine-Hugoniot relation \cite{Su2007}, displayed as solid circular dots. 
The comparison shows that these two Hugoniots agree reasonably well with each other. 
So, there is no fundamental obstacle to apply the eFF method in the simulation of a highly non-equilibrium process. 
Note that the Hugoniot in Fig.~\ref{pic_hugoniot} is simulated with $m_e$ = 0.01 amu, whereas the result through the Rankine-Hugoniot relation is calculated with $m_e$ = 1 amu. These results show that reducing the mass of electrons does not change the equilibrium property of D$_2$ in the downstream region of the shock front.

Also displayed are typical experimental measurements conducted in recent years\cite{Holmes1995PR, Knudson2001PRL, Knudson2003PRL} together with results of the  path-integral Monte Carlo (PIMC) method \cite{Militzer2000PRL}. 
They show that the largest compression ratio of D$_2$ is $\sim$ 4.3, whereas the eFF gives a slightly overestimated prediction of $\sim$ 5.2. 
This deviation are associated with the underestimation of the dissociation energy in the eFF method (67.2 kcal /mol for the eFF method and 104.2 kcal/mol for the exact bonding energy) \cite{Su2007}, which makes the material easier to compress.

\section{Summary}\label{sec_4}

In summary, a systematic study of shock wave propagating in dense deuterium  is carried out with the eFF method. 
Several non-equilibrium features associated with the excitation of electrons near the shock front are displayed, which afford a more complicated shock wave structure compared with the structure revealed by the methods that do not consider the effect of electrons.
The physical picture provided by the simulation could be helpful  to build a more realistic picture of shock wave propagation in fuel materials commonly used in ICF.

\begin{acknowledgments}
This work is financially supported by the NSFC (Grant No. 11274019) and NSAF (Grant No. U1530113).
Part of the calculations were supported by the Special Program for Applied Research on Super Computation of the NSFC-Guangdong Joint Fund (the second phase).
\end{acknowledgments}

\bibliography{eff_shock_mini}

\end{document}